 \documentclass[%
 aip,
 prl,%
 amsmath,amssymb,
preprint,%
 reprint,%
twocolumn,
]{revtex4-2}

\usepackage{graphicx}
\usepackage{dcolumn}
\usepackage{bm}
\usepackage{xcolor}

\begin{document}

\preprint{AIP/123-QED}

\title{A simple method for detection and quantitative estimation of deep levels in a barrier layer of AlGaN/GaN HEMT structures by analysis of light induced threshold voltage shift}

\author{Maciej Matys}
 \altaffiliation{Fujitsu Limited, Atsugi, Kanagawa, 243-0197, Japan}
 \email{matys.maciej@fujitsu.com}
\author{Atsushi Yamada}
\affiliation{Fujitsu Limited, Atsugi, Kanagawa, 243-0197, Japan}
\author{Yoichi Kamada}
\affiliation{Fujitsu Limited, Atsugi, Kanagawa, 243-0197, Japan}
\author{Toshihiro Ohki}
\affiliation{Fujitsu Limited, Atsugi, Kanagawa, 243-0197, Japan}

\date{\today}

\begin{abstract}
The characterization of deep levels in AlGaN/GaN heterostructures is one of the most important problems in GaN high electron mobility transistors (HEMTs) technology. This work reports on a technique for determination of deep level concentration in AlGaN/GaN HEMT structures. The proposed method is relatively simple, and it is based on the detection of free holes created by optically induced transitions of electrons from the deep levels to the conduction band. The developed method can detect and provide quantitative estimation of deep level traps in a barrier layer of AlGaN/GaN HEMT structures. Furthermore, it provides a framework for analysis of light induced threshold voltage shift, which includes an important experimental criterion of determination whether the holes are generated or not in the AlGaN/GaN HEMT structures by sub-band gap illumination. The method was verified by applications it to a study of the deep levels in GaN HEMTs grown on various substrates, i.e. SiC and GaN.
\end{abstract}
\maketitle

\section{Introduction}

Gallium nitride (GaN) high-electron mobility transistors (HEMTs) demonstrate outstanding performance in next-generation high-voltage and power applications\cite{KA,KB,K1,K2,K3,K4,K6a,K6a1,K6b,K6c}. However, despite the impressive progress which was made in improvement of performance of these devices, the reliability issues related to the charge trapping phenomena are still a challenging problem\cite{K5}. The typical effect of charge trapping phenomena is called "current collapse", i.e. a decrease of the drain current under the large drain voltage operation\cite{K6,K7}. According to several studies, this undesirable effect is caused by the electron trapping at the deep levels located in the buffer layers or/and at the surface\cite{K8,K9,K10,K11,K12}. Overall, the deep levels in GaN based HEMTs are present due to epitaxial growth of the HEMT structure\cite{K6,K7,K8,K13,K14} and process condition (for example surface etching\cite{K15,K16,K17}) or they can be induced by electrical stress\cite{K6}. In order to control the deep levels in GaN based HEMTs, the characterization of their properties is one of the fundamental aspects in the study of the GaN HEMTs. For quantitative analysis of the deep levels in GaN based devices the deep-level transient spectroscopy (DLTS)\cite{K17a} and deep-level optical spectroscopy (DLOS)\cite{K17b} are mainly applied\cite{K18}. These methods can provide information about the basic properties of the deep levels like their concentration, activation energy or capture cross sections. However, they have one important limitation, namely they cannot clearly indicate the spatial location of deep levels in the AlGaN/GaN HEMT structure\cite{K6}.

\begin{figure}
\includegraphics{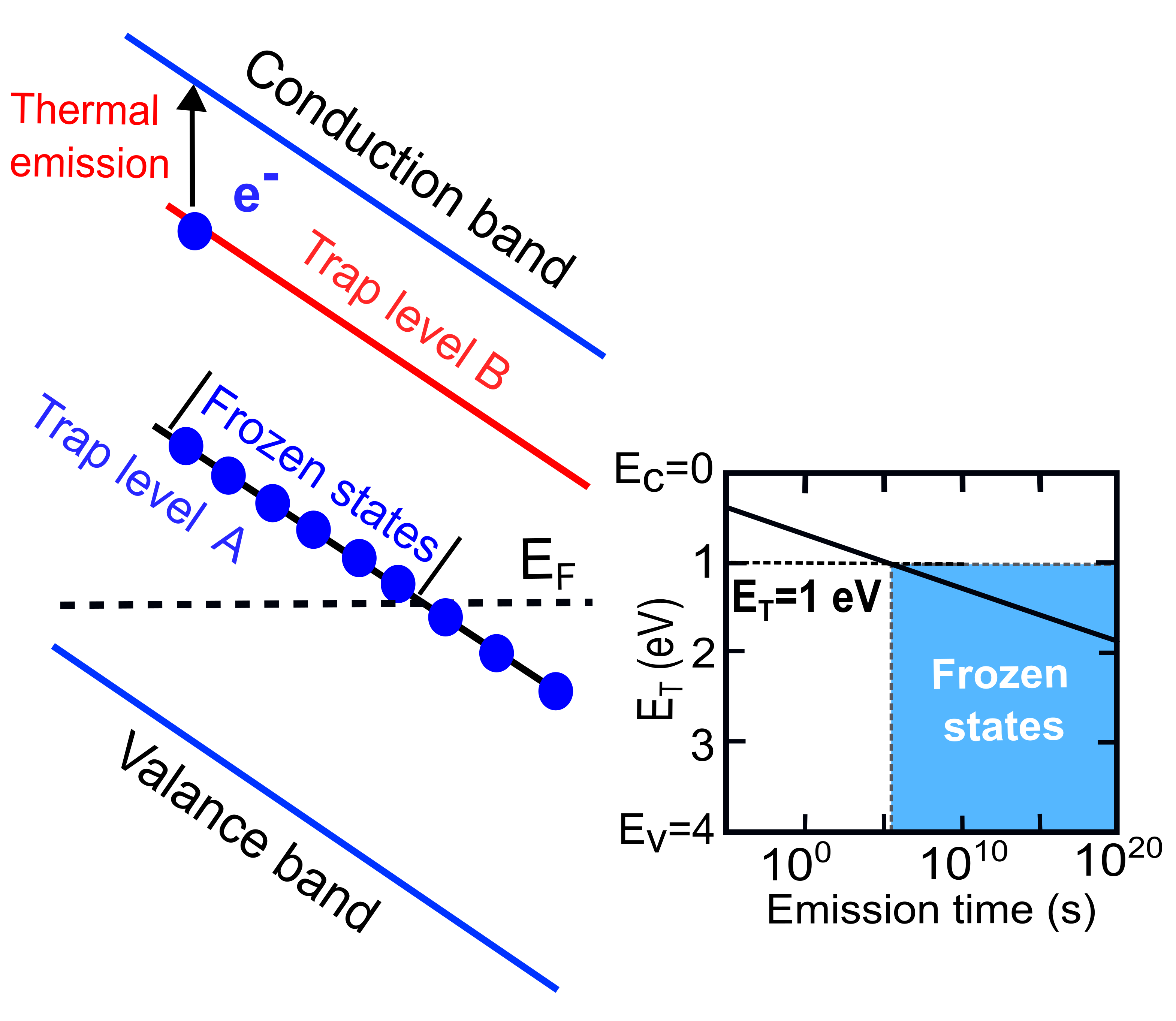}
\caption{\label{fig:epsart}Band diagram of the semiconductor device under reverse bias polarization with marked two trap levels A and B. Inset shows the calculated emission time of an electron from a trap with energy $E_T$ at RT in the case of Al$_{0.3}$Ga$_{0.7}$N. }
\end{figure}

In this paper, we present a simple method which can provide both (\emph{I}) quantitative estimation of deep-levels and (\emph{II}) precise information on their spatial location in the GaN HEMT structure. The key point of the proposed method is detection of free holes created by optically induced transitions of electrons from the deep levels to the conduction band (CB). We verified the developed technique by applications it to a study of the deep levels in GaN HEMTs grown on various substrates, i.e. SiC and GaN. The paper is organized as follows. In Section II, we describe the theoretical backgrounds of the proposed method. In Section III, we present the experimental results on deep levels obtained from GaN HEMTs on SiC and GaN substrates and in Section IV, we summarize the key points of this paper.

\section{Theory and background }

\subsection{Definition of deep level}

At first, for the future discussion, we explain on what basis to determine a defect state in the band gap as a deep one in a semiconductor device. Fig. 1(a) shows schematically the band diagram of an electronic device under the reverse bias polarization with marked two traps with energy levels A and B. For the considered reverse bias, we assumed that the position of the Fermi level ($E_F$) is below the trap level A (see Fig. 1(a)). The trap A is located at the larger energy distance much further from the conduction band (CB) compared to the trap B. Due to this fact, trap A is fully occupied by electrons while the trap B is empty from electrons. In other words, shifting Ef by the reverse bias below the energy level of trap A will not lead to electron emission from trap A in reasonable time in contrast to trap B for which emission electrons occurs almost immediately after shifting $E_F$ by the gate bias ($V_G$). Due to this difference, we can call the trap A as "frozen" or "deep". Fig. 1(b) shows the calculated emission time of electron ($\tau$) from a trap with energy $E_T$ at room temperature (RT) in the case of the AlGaN material, which is the main subject of this paper. The $\tau$ was calculated from the Shockley–Read–Hall (SRH) statistics according to the following equation\cite{K19}:

\begin{equation}
\tau=\frac{1}{v_{th}N_C\sigma}\exp(\frac{E_T}{kT})
\end{equation}

where $v_{th}=10{^6}$ cm/s, $N_C=10^{19}$ cm$^{-3}$, $\sigma=10^{-14}$ cm$^{-2}$ and $T$ are the electron thermal velocity, the density of states at CB, the capture cross section and temperature, respectively\cite{K20}. From Fig. 1(b) one can note that for the trap located at 1 eV $\tau$ is approximately $10^{5}$ s ($\approx$ 28 hours). Thus, for the AlGaN material (and also GaN) it can be safely assumed that every trap located deeper that 1 eV is "frozen" at RT, i.e. it does not change the electronic state under shifting $E_F$. This is important point to understand the shift of the threshold voltage ($V_{th}$) of the AlGaN/GaN Schottky barrier diode (SBD) upon light illumination, which will be discussed in the next section.

\subsection{Light induced $V_{th}$ shift of AlGaN/GaN SBD and hole emission}

\begin{figure}
\includegraphics{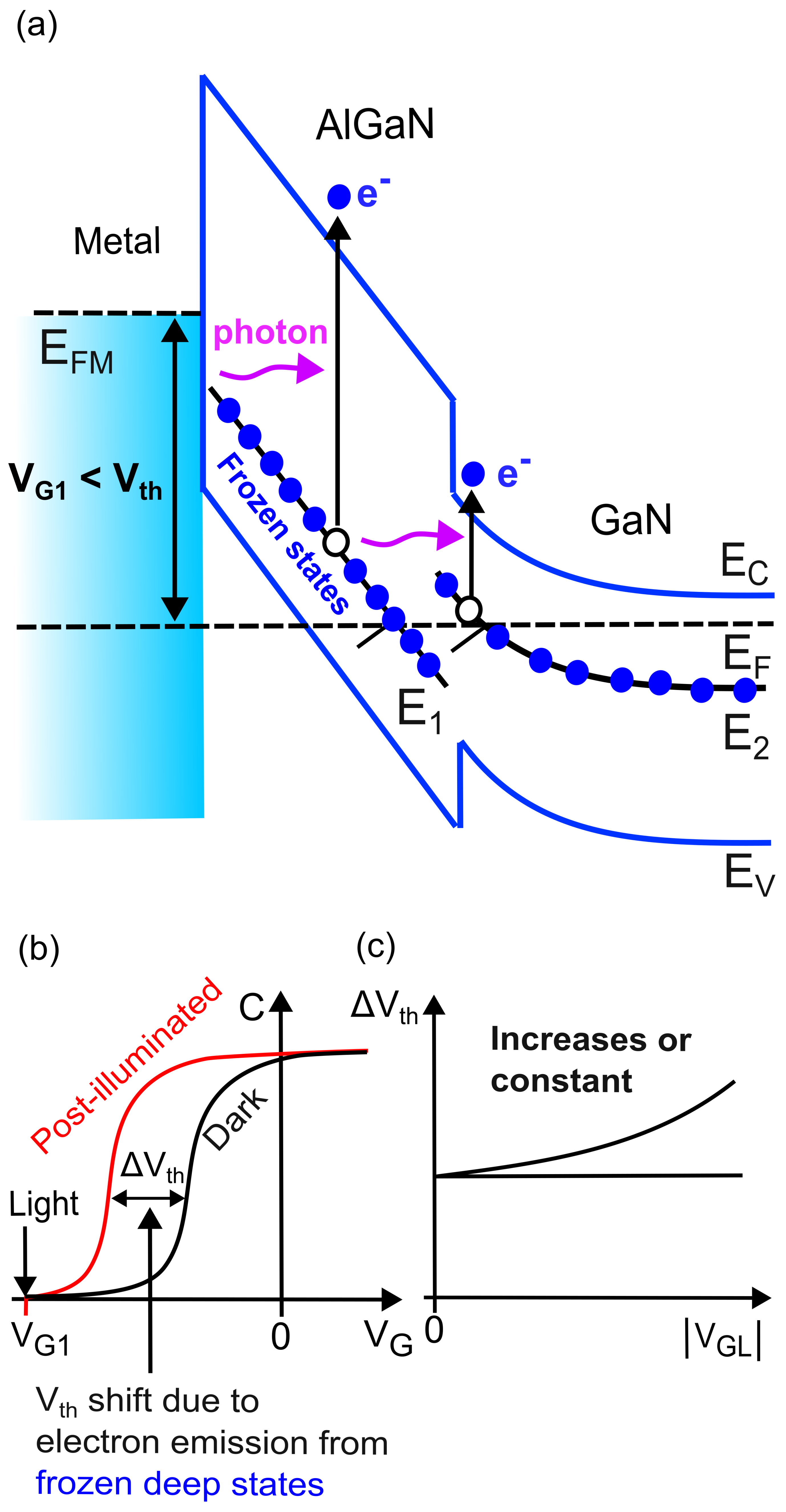}
\caption{\label{fig:epsart} (a) Band diagram of AlGaN/GaN SBD polarized under $V_{G1} < V_{th}$ in the case when only electron emission from the deep-levels $E_1$ and $E_2$ takes place. (b) Schematic illustration of light induced $V_{th}$ shift and (c) expected dependencies of $\Delta V_{th}$ with $V_{GL}$ when only electron emission from $E_1$ and $E_2$ occurs. }
\end{figure}

\begin{figure}
\includegraphics{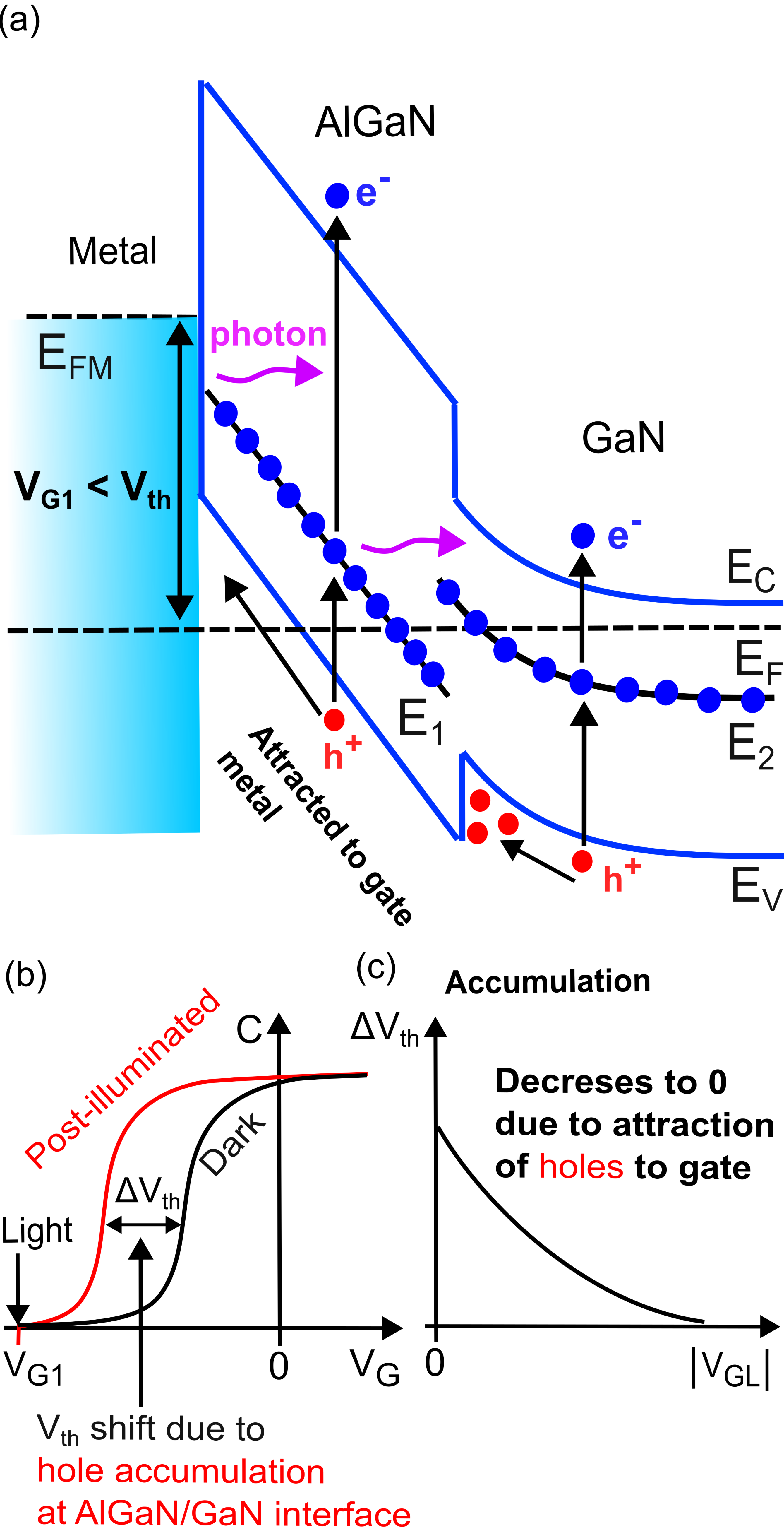}
\caption{\label{fig:epsart} (a) Band diagram of AlGaN/GaN SBD polarized under $V_{G1} < V_{th}$ in the case when both electron and hole emission from the deep-levels $E_1$ and $E_2$ takes place. (b) Schematic illustration of light induced $V_{th}$ shift and (c) expected dependencies of $\Delta V_{th}$ with $V_{GL}$ when both electron and hole emission via $E_1$ and $E_2$ levels occurs.}
\end{figure}

Let us consider now the AlGaN/GaN SBD diode polarized under the gate bias ($V_{G1}$) below $V_{th}$ whose band diagram is shown in Fig. 2(a). We assumed that both AlGaN and GaN layers contain some deep levels with energy $E_1$ and $E_2$ respectively. The deep levels $E_1$ and $E_2$ remain filled with electrons even though $E_F$ is located below these levels, which is in accordance with the previous definition of the deep level (Fig. 1). Next, the AlGaN/GaN SBD diode is illuminated by light with energy below the GaN band gap, which causes excitation of electrons from deep levels in AlGaN and GaN layer to CB, as shown in Fig. 2(a). Excitation of electrons from the deep levels $E_1$ and $E_2$ leads to $V_{th}$ shift with respect to it dark value when $V_G$ is sweeping (after turn of the light) from $V_{G1}$ to positive values, as shown schematically in Fig. 2(b). Now, we repeat all above processes (using the same light energy) i.e. $V_G$ is shifted firstly from the accumulation value to the gate bias below $V_{th}$ and then light is switched on. However, in this case $V_G$ at which illumination occurs is different than previously, for example more negative ($V_{G2}<V_{G1}$). After illumination at $V_{G2}$, the gate bias is sweeping from $V_{G2}$ to the positive values and a new $V_{th}$ shift is obtained. In this simple approach, which assumes only electron emission from the deep levels $E_1$ and $E_2$, the $V_{th}$ shift obtained for $V_{G2}$ and $V_{G1}$ should be similar or slightly higher for $V_{G2}$ due to increasing number of "frozen" deep states in the GaN layer (as a result of shifting $E_F$ in the GaN layer). In other words, when we consider only electron emission from the deep levels $E_1$ and $E_2$, the dependencies of $V_{th}$ shift ($\Delta V_{th}$) as a function of the gate bias at which illumination take place ($V_{GL}$) should be constant or slightly increased when $V_{GL}$ becomes more negative, as shown in Fig. 2(c).

The above picture becomes more complicated when holes are generated due to excitation of electrons from deep levels $E_1$ and $E_2$ to CB by light. Firstly, we consider the situation when for both deep levels $E_1$ and $E_2$ the hole emission takes place. After illumination of the AlGaN/GaN SBD diode polarized by $V_{G1}$ (below $V_{th}$) with light of energy below the GaN band gap, as previously, the deep levels $E_1$ and $E_2$ become empty from electrons. However, in this case, the electron transitions from the valence band (VB) to these empty levels take place leaving the holes in VB (see Fig. 3(a)). At this point, it is important to note that the occupation of deep-levels $E_1$ and $E_2$ does not change at all when the hole emission occurs (they are filled with electrons like before illumination, see Fig. 3(a)). Since the structure is polarized by the negative bias all free holes created in the AlGaN layer are attracted to the gate metal, as shown in Fig. 3(a). On the other hand, if $V_{G1}$ is not too strong (i.e. is not too negative) the holes generated in the GaN layer have a chance to accumulate at the AlGaN/GaN interface (because of the low electric field) instead of being attracted to the gate metal like free holes in the AlGaN layer (see Fig. 3(a)). Accumulated free holes at the AlGaN/GaN interface act as the additional positive charge leading to $V_{th}$ shift toward to the negative values when the $V_G$ is sweeping from $V_{G1}$ to the positive values (see Fig. 3(b)). Thus, as in the previous case (Fig. 2) we obtain a $V_{th}$ shift, however the origin of this shift is totally different. Now, if the gate bias at which illumination take place i.e. $V_{GL}$ will be more negative, the magnitude of $V_{th}$ shift should decrease since the free holes accumulated at the AlGaN/GaN interface (Fig. 3(a)) will be attracted to the gate metal by the strong electric field. In consequence, $\Delta V_{th}$ should be a decreasing function of $V_{GL}$ going to zero at the large negative $V_{GL}$, as schematically shown in Fig. 3(c), when the hole emission occurs for both levels $E_1$ and $E_2$.

\begin{figure}
\includegraphics{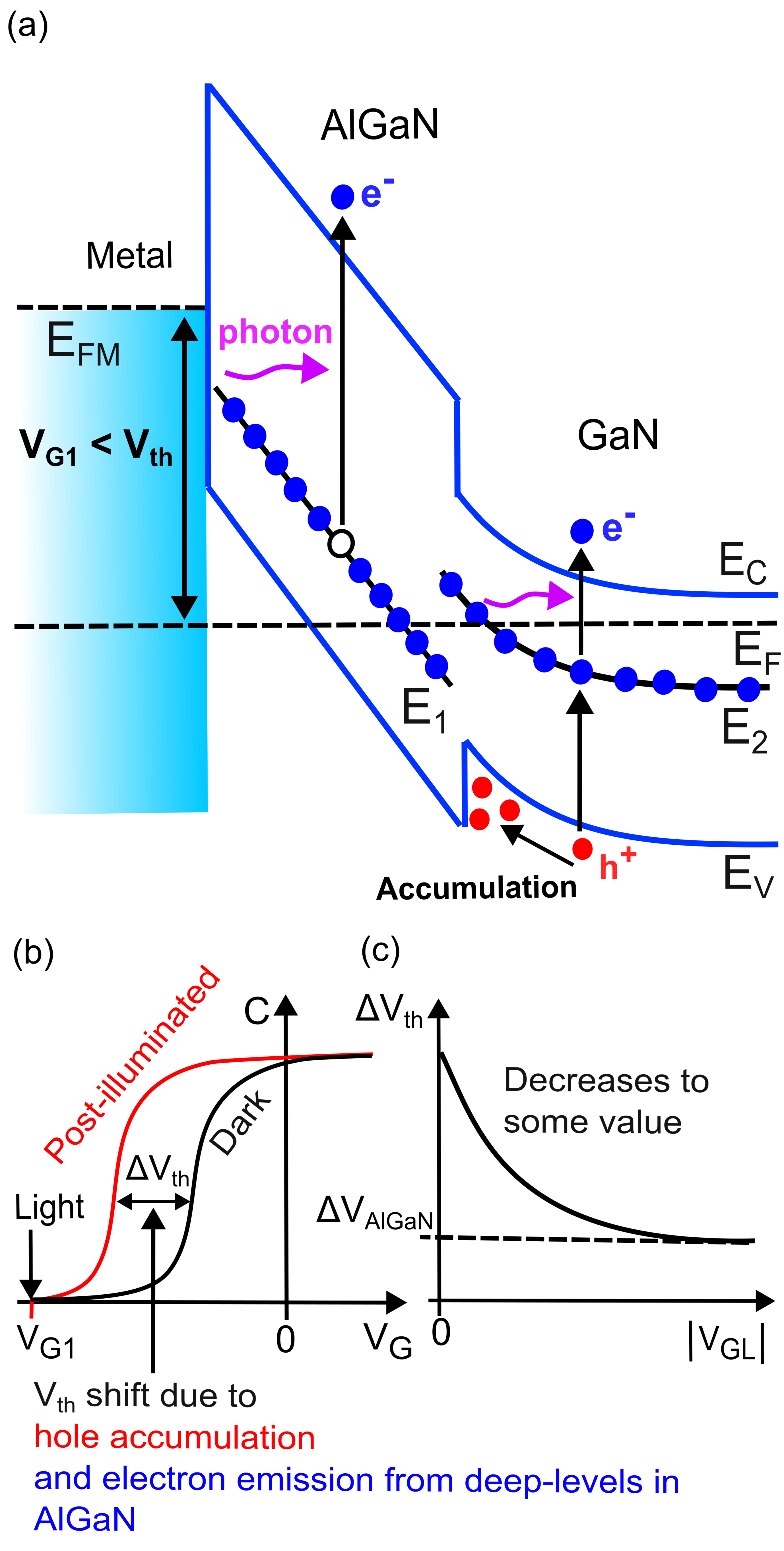}
\caption{\label{fig:epsart} (a) Band diagram of AlGaN/GaN SBD polarized under $V_{G1} < V_{th}$ when electron emission from $E_1$ and hole (and electron) emission via $E_2$ occur. (b) Schematic illustration of light induced $V_{th}$ shift and (c) expected dependencies of $\Delta V_{th}$ with $V_{GL}$.}
\end{figure}

The most interesting case is the situation when the hole emission occurs only in the GaN layer but not in AlGaN one. Fig. 4(a) shows the band diagram of the illuminated AlGaN/GaN SBD diode, polarized under $V_{G1}$ (below $V_{th}$) assuming the hole emission only via the $E_2$ level in the GaN layer (in the case of $E_1$ level in AlGaN layer, light excites only electrons like in Fig. 2(a)). As in the previous case, holes generated in the GaN layer due to excitation of electrons from the $E_2$ level are accumulated at the AlGaN/GaN interface. On the other hand, due to excitation of electrons, the $E_1$ level in AlGaN becomes empty from electrons, i.e. it changes the electronic state as shown in Fig. 4(a). Both these processes lead to $V_{th}$ shift when the gate bias is sweeping from $V_{G1}$ toward accumulation (Fig. 4(b)). Thus, compared to the previous case (Fig. 3), the $V_{th}$ shift is not only related to holes in the GaN layer but also to the deep levels in the AlGaN layer. In this scenario, when $V_{GL}$ becomes more negative, the magnitude of $V_{th}$ shift should firstly decrease due to attraction of the accumulated holes at the AlGaN/GaN interface to the metal gate and subsequently becomes a constant with $V_{GL}$, as shown in Fig. 4 (c). The constant value ($\Delta V_{AlGaN}$, see Fig. 4(c)) reached after decreasing $\Delta V_{th}$ with $V_{GL}$ is purely related to the deep levels in AlGaN layer. Thus, from $\Delta V_{AlGaN}$ the concentration of deep levels in AlGaN layer ($N_{Deep}$) can be roughly estimated according to the following equation (see Appendix A):

\begin{equation}
N_{Deep}=\frac{2C_{total}\Delta V_{AlGaN}}{qt_{AlGaN}}
\end{equation}

where $C_{total}$ is the AlGaN layer capacitance and $t_{AlGaN}$ is the AlGaN layer thickness. In the opposite situation, i.e. when the hole emission occurs only in the AlGaN layer but not in GaN one, the dependencies of $V_{th}$ shift as a function of $V_{GL}$ should be similar as in the Fig. 2(c) since all the generated holes in AlGaN layer are attracted to the gate metal. As a consequence, after the illumination only the deep levels in GaN layer change their occupation that results in $V_{th}$ shift. Thus, an increase of $V_{GL}$ should not be caused by a decrease of $\Delta V_{th}$, like in Figs. 3 and 4. Instead of this $\Delta V_{th}$ should be constant or slightly increasing with $V_{GL}$ due to an enlargement of the amount of "frozen states" in the GaN layer.

\begin{figure}
\includegraphics{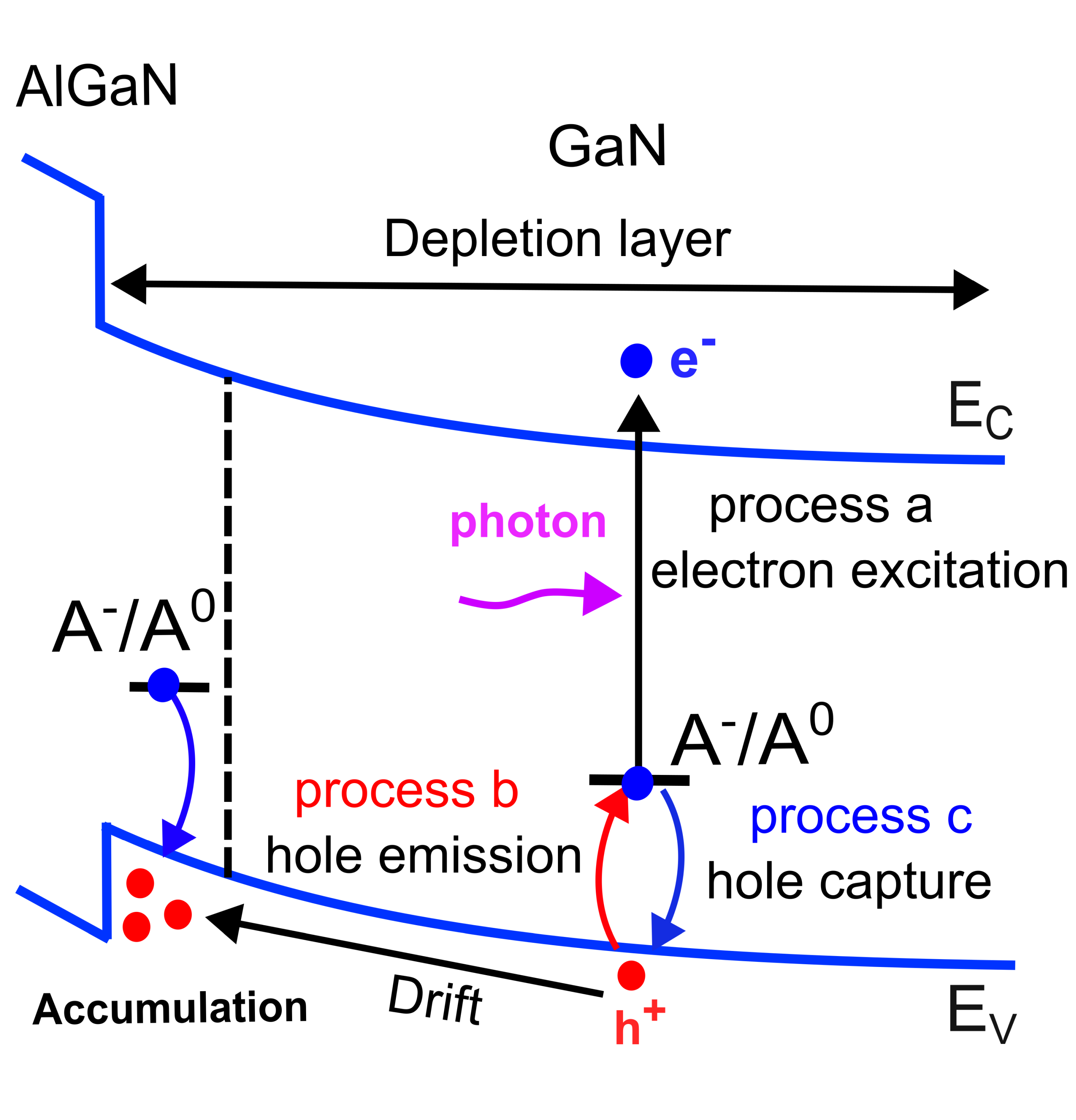}
\caption{\label{fig:epsart} Band diagram of GaN layer with the following processes marked: excitation of electron (by light) from a deep-acceptor state ($A^{0}$/$A^{-}$) to CB (process a), hole emission, i.e. transition of electron from VB to deep-acceptor state (process b) and hole capture (transition of electron from deep-acceptor state to empty space (hole) in VB.}
\end{figure}

\subsection{Hole capture process}

It should be added that in the discussion in subsection B, we omitted some possible electron transition, i.e. capturing of photo-induced free holes by an ionized deep level. This is because such process has a negligible impact for analysis of the $V_{th}$ shift of C-V curves, as is proved below based on the detailed model of the electronic transitions.

In general, our model considers the hole generation in VB due to optically induced transitions of electrons from the deep levels to CB. In detail, this process was schematically illustrated in Fig. 5 which shows the band diagram of GaN containing a deep-acceptor state ($A^{0}$/$A^{-}$). Due to sub-band gap illumination, electron from the deep acceptor state is excited to CB leading to a change of the acceptor state from $A^{-}$ to $A^{0}$, as shown in Fig. 5 (process a). After this process, the deep acceptor state is empty from electron (state $A^{0}$) and thus electron from VB can jump to this empty state leaving the hole in VB (process b). As a consequence of process b, the acceptor state changes again the state from $A^{0}$ to $A^{-}$. Compared to the previous discussion in the subsection B, at this point we included an additional possible electronic transition, i.e. capturing of a free hole generated in VB (due to process b) by an ionized acceptor state ($A^{-}$) (process c, Fig. 5) leading to the change in the acceptor state from $A^{-}$ to $A^{0}$. However, as we explain, the capture process c can be completely ignored in the analysis of the $V_{th}$ shift of C-V curves. 

At first, it should be noted that all considered processes occur in the depletion regions (in our experiment, we measured the capacitance changes) which means that the generated holes in VB due to process b are quickly swept by the strong electric field toward the AlGaN/GaN interface (see Fig. 5) and toward the gate metal (in the case of AlGaN layer, see Fig. 3(a)). For example, in the typical GaN depletion electric field, i.e. 10$^{5}$ V/cm, the carriers are separated within the time interval of 10$^{-11}$s\cite{K20a}, whereas the capture time of holes by the ionized acceptor state ($A^{-}$) is of the order of 10$^{-4}$-10$^{-7}$s\cite{K20b}. This clearly indicates that just after generation, the free holes move immediately toward the AlGaN/GaN interface where they are accumulated in a thin region, as shown in Fig. 5. As a consequence, only the deep acceptor state located in a thin region around AlGaN/GaN interface can capture free holes and change the electronic state from $A^{-}$ to $A^{0}$ (see Fig. 5). In other words, the process c is important only in a small part of the depletion region (near the AlGaN/GaN interface), whereas in the remaining dominant part of depletion region this process can be totally ignored. This translates into the fact that the total number of captured holes (due to process c) should be negligible compared to the total number of generated holes in the whole depletion region. Therefore, the process c can be omitted in the analysis of the $V_{th}$ shift of C-V curves.

\begin{figure}
\includegraphics{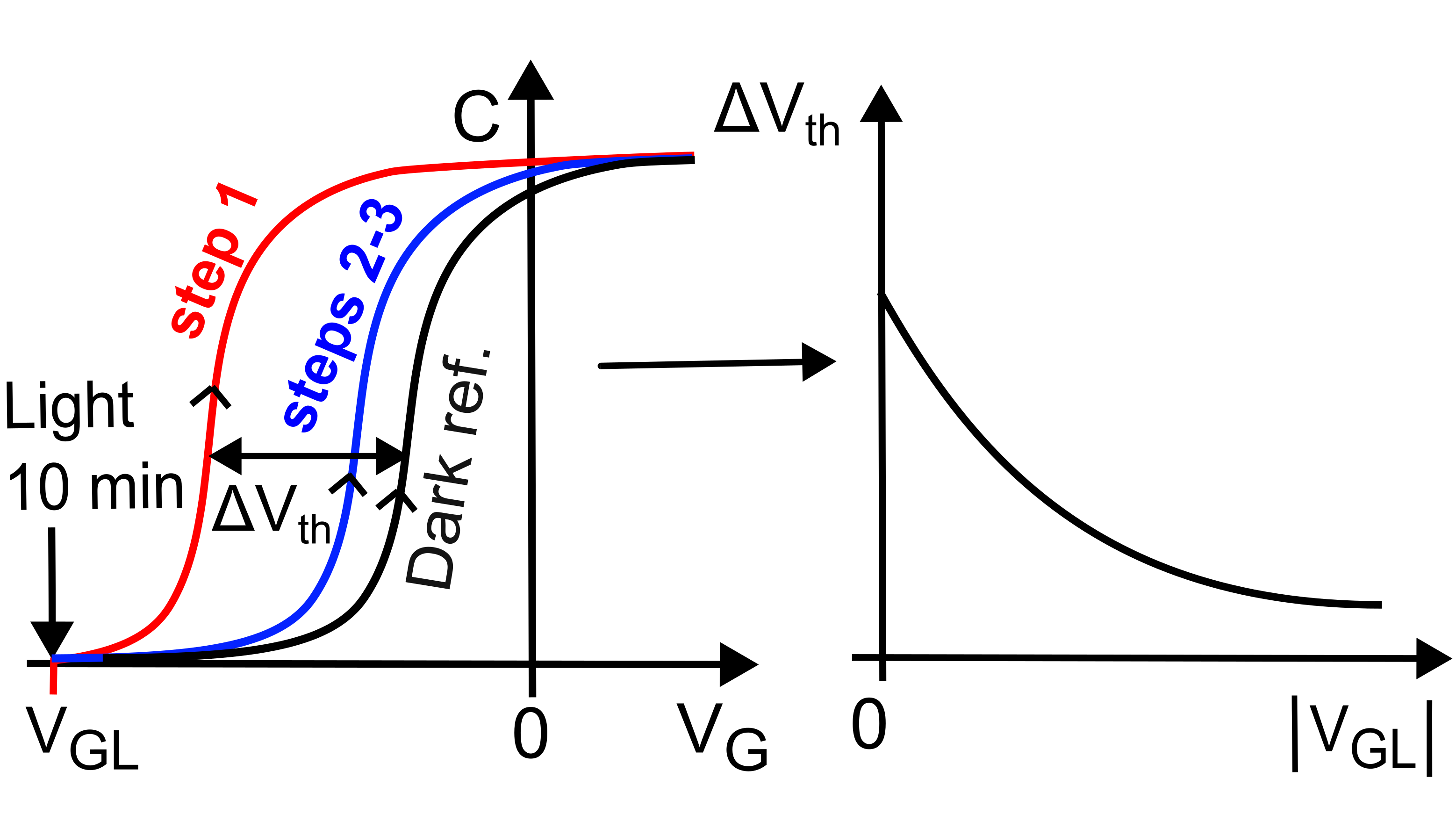}
\caption{\label{fig:epsart} Procedure for estimation of deep-level concentration in AlGaN barrier.}
\end{figure}

\subsection{Procedure for estimation of deep-level concentration in AlGaN barrier}

Based on the above consideration, we propose the following procedure for the estimation of the concentration of deep levels in the AlGaN barrier of the AlGaN/GaN heterostructure. Firstly, $V_G$ of AlGaN/GaN SBD is sweeping from the depletion to positive (accumulation) bias to fill all deep levels with electrons (Fig. 6, step 1). Next, $V_G$ is sweeping from the positive bias to $V_{GL}$ below $V_{th}$ (step 2). Subsequently, keeping the bias at $V_{GL}$, the AlGaN/GaN SBD diode is illuminated using the sub-band gap light during 10 min. Such long illumination time was chosen to ensure the full depopulation of deep levels from electrons. After the light-off, $V_G$ is sweeping from $V_{GL}$ to accumulation (Fig. 6, step 3) and the $V_{th}$ shift is registered. All steps from 1 to 3 are repeated for different $V_{GL}$ values (all below $V_{th}$) and the dependencies of $\Delta V_{th}$ as a function $V_{GL}$ are obtained, as shown in Fig. 6. Next, from these dependencies one can deduce, which scenario occurs (Figs. 2-4). In particular, if $\Delta V_{th}$ decreases with $V_{GL}$ down to a constant value (see $\Delta V_{AlGaN}$ in Fig. 4(c)) the hole generation occurs only in the GaN layer (see Fig. 4). In this case, the determining constant value of $\Delta V_{AlGaN}$ reached after decreasing $\Delta V_{th}$ with $V_{GL}$ and using Eq. 2 allow the calculation of $N_{Deep}$.

\subsection{Application to other structures}

\begin{figure}
\includegraphics{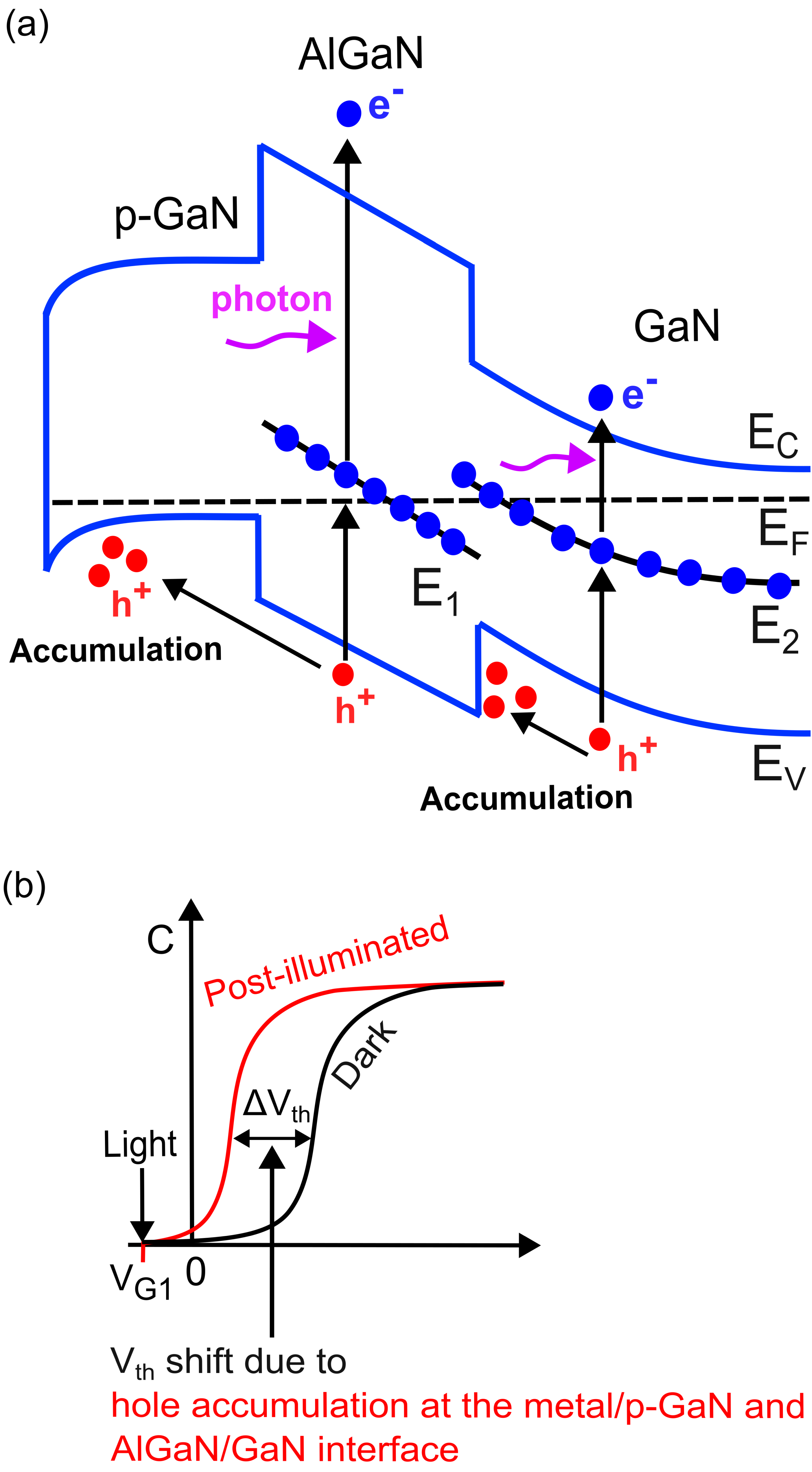}
\caption{\label{fig:epsart}(a) Band diagram of metal/p-GaN/AlGaN/GaN structure polarized under $V_{G1} < V_{th}$ and illuminated by the light with energy lower than the GaN band gap. (b) Schematic illustration of light induced $V_{th}$ shift in the metal/p-GaN/AlGaN/GaN structure.}
\end{figure}

So far, we have considered only the metal/AlGaN/GaN HEMT structures. However, it is interesting to comment briefly if the proposed method can be applied also to other structures, for example p-GaN/AlGaN/GaN HEMT structures. The main problem of the application of the developed method to p-GaN/AlGaN/GaN HEMT structures lies in the top p-GaN layer. Due to the presence of this layer, it is practically impossible to separate signal from the AlGaN and GaN buffer layers using the described approach which means that the proposed method cannot be effective in such structures. In more details, this problem can be explained as follows.

Fig. 7(a) shows the band diagram of the metal/p-GaN/AlGaN/GaN HEMT structure polarized by the gate bias ($V_{G1}$) below the threshold voltage. Under such conditions all free holes created in the AlGaN layer due to sub-band gap illumination are attracted to the gate metal, as shown in Fig. 7(a). However, because of the p-GaN layer they are accumulated near the metal/p-GaN interface instead of flowing to the gate metal as in the case of a simple metal/AlGaN/GaN structure (see Fig. 3(a)). On the other hand, if $V_{G1}$ is not too negative, the free holes generated in the GaN layer are accumulated at the AlGaN/GaN interface (Fig. 7(a)) (instead of being attracted toward the gate metal). The accumulated free holes at the metal/p-GaN and AlGaN/GaN interfaces act as the additional positive fixed charge leading to the $V_{th}$ shift (when the gate bias is sweeping from $V_{G1}$ to the positive values, see Fig. 7(b)). This means that in the case of the p-GaN/AlGaN/GaN HEMT structures it is practically impossible to determine which free holes (from GaN or AlGaN layer) cussed the $V_{th}$ shift after sub-band gap illumination. Similar situation occurs also in the metal-insulator AlGaN/GaN structures. However, in such structures free holes from the AlGaN layer will accumulate at the insulator/AlGaN interface (due to the presence of the valence band offset) instead of flowing to the gate metal. Therefore, only in the Schottky based AlGaN/GaN HEMT structures we can clearly distinguish which holes lead to the $V_{th}$ shift since in these structures free holes generated in the AlGaN layer due to sub-band gap illumination can directly flow to the gate metal (see Fig. 3(a)).

\begin{figure}
\includegraphics{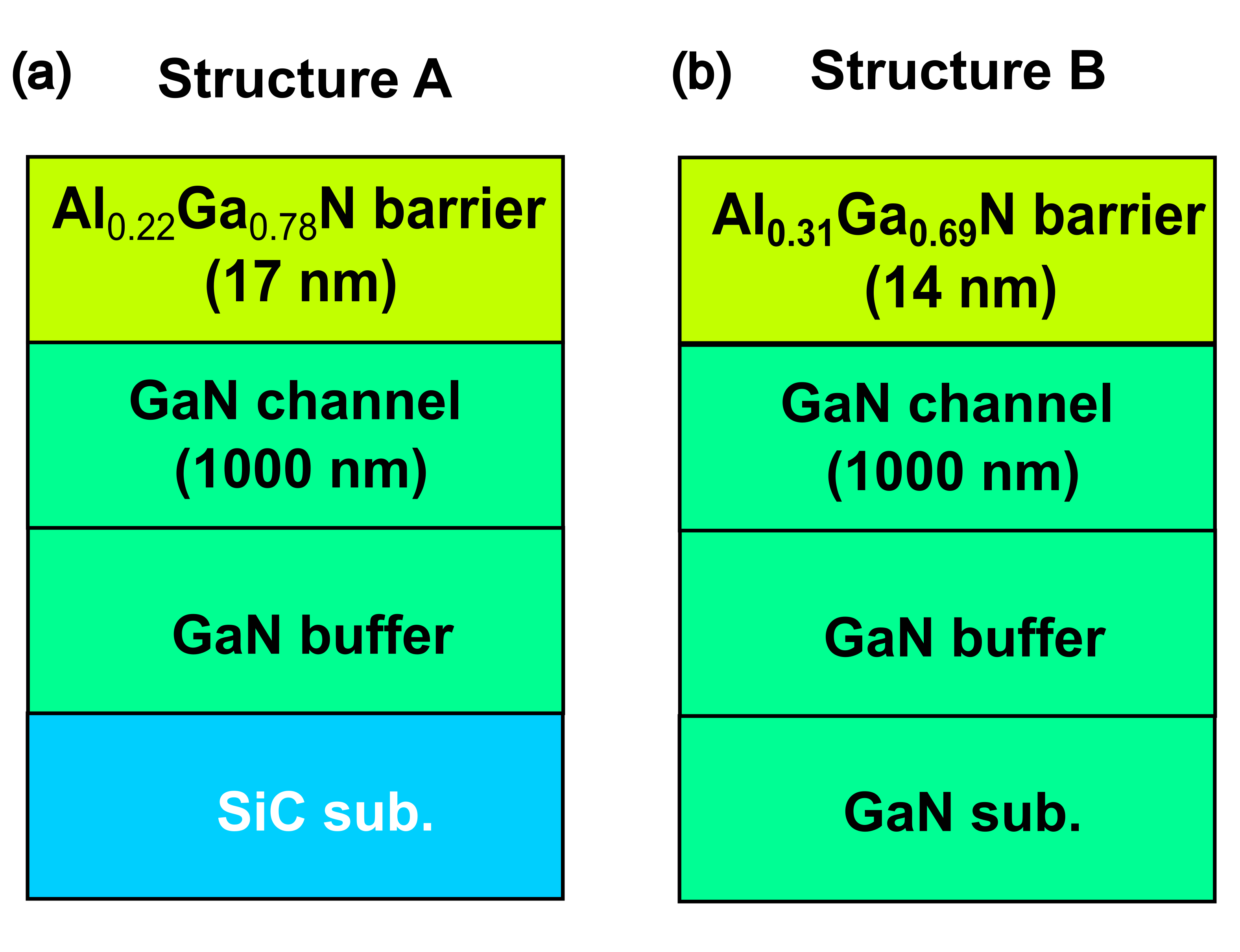}
\caption{\label{fig:epsart} Schematic illustration of the epitaxial structures used in this study.}
\end{figure}
 
\section{Results and discussion}

\subsection{Epitaxial structures and experimental conditions}

Fig. 8 shows the schematic illustration of the AlGaN/GaN HEMT structures used in this study. The structure A was grown on a SiC substrate (Fig. 8(a)) while the structure B was grown on a GaN substrate (Fig. 8(b)). Epitaxial growth of the structure A was performed using a horizontal flow metalorganic vapor phase epitaxy (MOVPE) reactor. The epitaxial structure A comprised a 14-nm Al$_{0.22}$Ga$_{0.78}$N barrier, a 1000-nm GaN channel, and a Fe-doped GaN buffer on a 3-inch semi-insulating SiC substrate (see Fig. 8(a)). For the growth of epitaxial structure B a semi-insulating, Fe-doped GaN crystal grown by hydride vapor phase epitaxy (HVPE) was used as a substrate. The epitaxial structure B comprised 2000 nm Fe-doped GaN buffer layer, 1000 nm GaN channel layer and 14 nm Al$_{0.31}$Ga$_{0.69}$N barrier layer (Fig. 8(b)). The details of the sample B fabrication process were reported in Ref. [31]. The photo-assisted capacitance-voltage (C-V) characteristics were obtained at 1MHz using an impedance analyzer. As a light source, 150 W halogen lamp with band-pass filters was applied. All measurements were performed at room temperature.

\subsection{Photo-assisted C-V characteristics}

\begin{figure}
\includegraphics{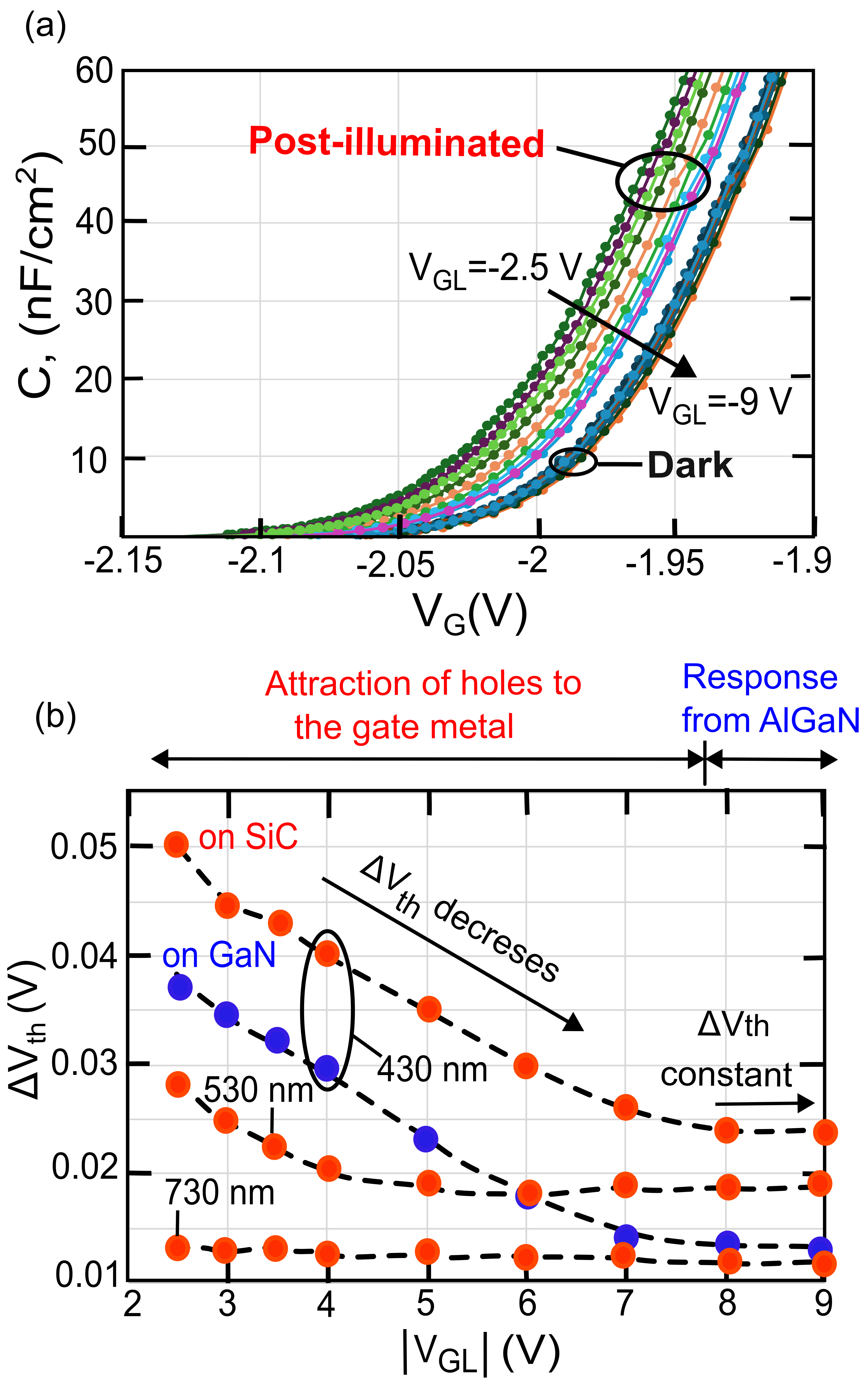}
\caption{\label{fig:epsart}(a) Photo-assisted C-V characteristics of AlGaN/GaN SBD structure A after illumination with 430 nm wavelength at various $V_{GL}$ from -9 V to -2.5 V. (b) Dependencies of $\Delta V_{th}$ as a function of $V_{GL}$ for structure A (wavelengths of 430 nm, 530 nm and 730 nm) and structure B (430 nm).}
\end{figure}

Fig. 9(a) shows the photo-assisted C-V characteristics of AlGaN/GaN SBD structure A after illumination with a wavelength of 430 nm at various $V_{GL}$ from -9 V to -2.5 V. One can note that the dark C-V curve (after 10 minutes of holding $V_{GL}$) practically did not change the position with $V_{GL}$. However, the C-V curve after 10 min illumination clearly changes the position with $V_{GL}$. In particular, for $V_{GL}$=-2.5 V, the post illuminated C-V curve is mostly shifted toward the negative $V_G$ values with respect to the dark C-V one. On the other hand, when $V_{GL}$=-9 V the post illuminated C-V curve is the closet to the dark C-V one. In Fig. 9(b), we summarized the dependencies of $\Delta V_{th}$ from Fig. 9(a) as a function of $V_{GL}$. In addition, in the same figure we also showed the dependencies of $\Delta V_{th}$ with $V_{GL}$ obtained for the wavelengths of 530 nm and 730 nm. As can be seen, for the 430 nm and 530 nm wavelengths $\Delta V_{th}$ clearly decreases with $V_{GL}$ up to a certain value but for 730 nm it is nearly constant with $V_{GL}$. Based on the theory provided in Sec. II, one can conclude that in the case of the sub-band gap illumination with wavelengths of 430 nm and 530 nm, the hole generation in the GaN layer occurs while for the 730 nm wavelength only electron emission takes place. In particular, a decrease of $\Delta V_{th}$ with $V_{GL}$ for 430 nm and 530 nm wavelengths is due to attraction of the holes accumulated at the AlGaN/GaN interface to the gate. For the comparison, in Fig. 9(b) we also showed the dependencies of $\Delta V_{th}$ as a function of $V_{GL}$ for the structure B obtained using the 430 nm wavelength. In this case $\Delta V_{th}$ also clearly decreases with $V_{GL}$ down to the constant value which is much smaller than in the case of structure A. As we explained in Sec. II, the constant value to which $\Delta V_{th}$ decreases is directly related to deep levels in the AlGaN layer. For the structure A, $\Delta V_{AlGaN}$=0.024 V while for the structure B $\Delta V_{AlGaN}$=0.012 V at the 430 nm wavelength illumination (see Fig. 9(b)). Using these values, from Eq. 2 we estimated the following concentration of deep-levels in the AlGaN barrier: for the structure A $N_{deep}$=9.5$\times10^{16}$ cm$^{-3}$ and for the structure B $N_{deep}$=6$\times10^{16}$ cm$^{-3}$ (assuming experimental $C_{total}$=542 nF/cm${^2}$ and $C_{total}$=571 nF/cm${^2}$ for structure A and B respectively). These results seem to be reasonable since the structure B is grown on the GaN substrates and thus it is expected that the AlGaN barrier quality is better than in structure A. Furthermore, the estimated concentrations are in a good accordance with the previous studies by Amstrong et al.\cite{K22} of deep levels near VB in the AlGaN layer.

\subsection{Pulsed current-voltage measurement}

\begin{figure}
\includegraphics{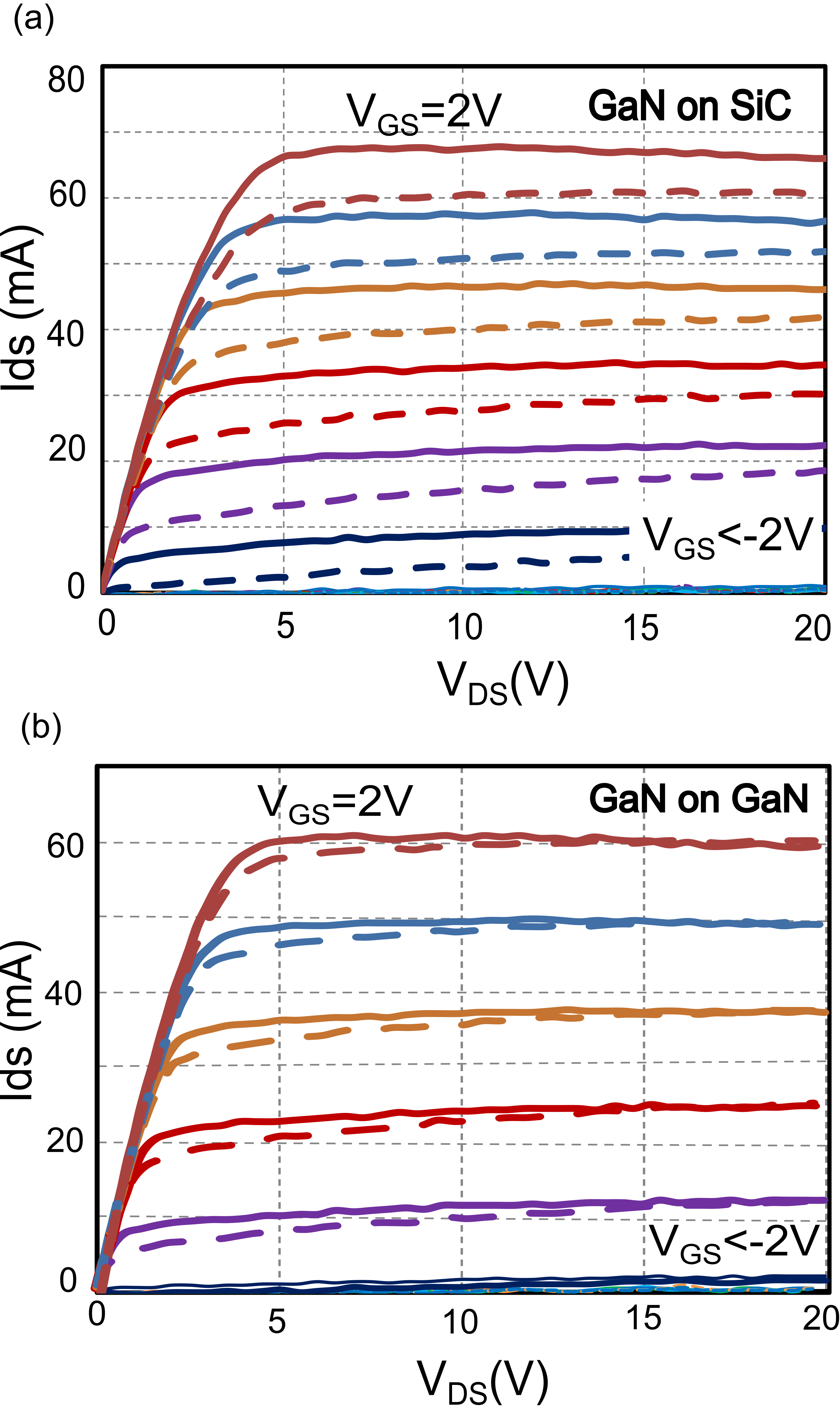}
\caption{\label{fig:epsart}Pulsed current-voltage characteristics of (a) structure A and (b) structure B. The gate voltage, $V_{GS}$, was changed by the step 0.5 V.}
\end{figure}

Finally, to evaluate independently the quality of the investigated epi layer HEMT structures, we performed the pulsed current (I)-voltage (V) measurement, as shown in Fig. 10. The pulsed I–V measurements were performed under the quiescent gate and drain voltage ($V_{Gsq}$, $V_{Dsq}$) of (0 V, 0 V) "without off-stress" and (–5 V, 50 V) "with off-stress". In the case of structure A, reduction of the drain current ($I_{ds}$) around the knee voltage by applying off-stress was observed (Fig. 10(a)). However, in the case of structure B, the drain current is almost the same with and without off-stress (Fig. 10(b)).

In the early reports, the current collapse effect was attributed to the carbon related acceptors\cite{K22L,K22a,K22b} or heating effects\cite{K23a}. However, in our case, it is more likely that the observed current collapse (in structure A) is caused by the Fe related traps (for example, Fe$_{Ga}$ acceptor state) since the investigated HEMT structures were grown with Fe doped GaN buffer layers (see Fig. 8). The Fe related traps are located around 0.6 eV from CB as it was experimentally found by Horita et al\cite{K13} and Kruszewski et al\cite{K14}. Our method does not provide information about such traps; it mainly determines the traps located around the midgap (or VB). Nevertheless, from the pulsed I-V measurements characteristics (Fig. 10) we can clearly say that the structure B (on GaN) exhibits much better quality (lower trap concentration) than the structure A (on SiC) which is in accordance with the results obtained from our method (see Fig. 9).

\section{Summary}

We proposed a relatively simple method for determining the deep level concentration in AlGaN/GaN HEMT structures. The developed method can detect and provide quantitative estimation of deep levels in the barrier layer of AlGaN/GaN HEMT structures. The key point of the proposed method is detection of the free holes created by optically induced transitions of electrons from the deep levels to the conduction band. The developed method can detect and provide quantitative estimation of deep level traps in barrier layer of AlGaN/GaN HEMT structures. Furthermore, it provides an important experimental criterium of determination whether the holes are generated or not in the AlGaN/GaN HEMT structures by sub-band gap illumination. The method was well verified by its application to studying the deep levels in GaN HEMTs grown on various substrates, i.e. SiC and GaN.

\section{Appendix A: derivation of Equation 2}

The threshold voltage of AlGaN/GaN HEMT structure is given by\cite{K23}: 

\begin{multline}
V_{th}=\frac{\phi_b}{q}-\frac{t_{AlGaN}\sigma_p}{\epsilon}-\frac{\Delta E_C}{q}-\\
\frac{\Delta E_{f0}}{q}+\frac{q}{\epsilon}\int_{0}^{t_{AlGaN}}N_{Deep}(x)xdx
\end{multline}

where $\phi_b$, $\sigma_p$, $\Delta E_C$, $E_{f0}$ is the metal-semiconductor Schottky barrier height, the polarization charge at the AlGaN/GaN interface, the  conduction band offset at the AlGaN/GaN and the difference between the Fermi level and the conduction band edge of the GaN channel and $N_{Deep}(x)$ is the total concentration of ionized deep acceptors, respectively. 

Assuming that the deep acceptor state ($E_1$ level) in the AlGaN layer is distributed uniformly, i.e. $N_{Deep}(x)$=constant, we obtain the following equation:
 
\begin{multline}
V_{th}=\frac{\phi_b}{q}-\frac{t_{AlGaN}\sigma_p}{\epsilon}-\frac{\Delta E_C}{q}-\\
\frac{\Delta E_{f0}}{q}+\frac{q}{2\epsilon}N_{Deep}t_{AlGaN}^2
\end{multline}

Upon illumination, the $E_1$ level in the AlGaN layer changes its occupation (due to electron emission) i.e $E_1$ level becomes neutral (or positive if it is an donor state) since we assume that there is no hole transitions in this case (see Fig. 4(a)). On the other hand, in the GaN layer, the hole generation occurs via the $E_2$ level, as shown in Fig. 4(a) which means that this level does not change its electronic state. The holes generated in the GaN layer are accumulated at the AlGaN/GaN interface from where they are attracted to the metal gate if the gate is too negative. Thus, the magnitude of the $V_{th}$ shift firstly decreases (due to attraction of the accumulated holes to the metal gate) and subsequently becomes constant versus $V_{GL}$, as shown in Fig. 4 (c) and in experimental data (Fig. 9). The constant value of ($\Delta V_{AlGaN}$, see Fig. 4(c)) reached at the large $V_{GL}$ (in experiment this value is around 8 V and 9 V, see Fig. 9 (b)) is due to photoionization of deep levels in the AlGaN layer. In consequence, the $V_{th}$ shift in C-V curves obtained at large negative $V_{GL}$ can be expressed as:
 
\begin{equation}
V_{th}=\frac{\phi_b}{q}-\frac{t_{AlGaN}\sigma_p}{\epsilon}-\frac{\Delta E_C}{q}-\frac{\Delta E_{f0}}{q}
\end{equation}

The difference between Eq. 5 and 4 is $\Delta V_{AlGaN}$. Thus, from the combination of Eq. 5 and 4, we get the equation:
 
\begin{equation}
N_{Deep}=\frac{2C_{total}\Delta V_{AlGaN}}{qt_{AlGaN}}
\end{equation}

\begin{acknowledgments}
The authors expresses gratitude to Norikazu Nakamura for his kind support and discussions. 
\end{acknowledgments}

\textbf{DATA AVAILABILITY}

The data that support the findings of this study are available from the corresponding authors upon reasonable request.


\end{document}